\documentclass[aps,prl,twocolumn,superscriptaddress,amsfont,graphicx,nofootinbib,preprintnumbers]{revtex4}
\usepackage{amsmath}
\usepackage{amssymb}
\usepackage{graphicx}
\usepackage{booktabs}
\usepackage{hyperref}
\usepackage{subfigure}

\newcommand{\NLL}{\text{NLL}}
\newcommand{\NNLL}{\text{NNLL}}
\newcommand{\FNLL}{\mathcal{F}_{\rm NLL}}

\newcommand{\ycut}{y_{\rm cut}}
\newcommand{\ythree}{y_{3}}
\newcommand{\dZ}{d{\cal Z}[\{R'_{\mathrm{NLL}, \ell_i}, k_i\}]}

\newcommand{\yfun}[1]{y_3(\{\tilde p \},#1)}
\newcommand{\yscbar}[1]{\overline{y}_3^{\rm sc}(\{\tilde p \},#1)}
\newcommand{\ysc}[1]{y_3^{\rm sc}(\{\tilde p \},#1)}
\newcommand{\yhc}[1]{y_3^{\rm hc}(\{\tilde p \},#1)}
\newcommand{\ywa}[1]{y_3^{\rm wa}(\{\tilde p \},#1)}

\begin{document}

\title{The two-jet rate in $e^+e^-$ at next-to-next-to-leading-logarithmic order}

\preprint{CERN-TH-2016-149, OUTP-16-19P}

\def\oxford{Rudolf Peierls Centre for Theoretical Physics, University
  of Oxford OX1 3PN Oxford, United Kingdom}
\def\cern{CERN, Theoretical Physics Department, CH-1211 Geneva 23, Switzerland}
\def\sussex{Department of Physics and Astronomy, University of Sussex,
  Falmer, Brighton BN1 9RH, United Kingdom}

\author{Andrea~Banfi}
\affiliation{\sussex}
\author{Heather~McAslan}
\affiliation{\sussex}
\author{Pier~Francesco~Monni}
\affiliation{\oxford}
\author{Giulia~Zanderighi}
\affiliation{\oxford}
\affiliation{\cern}

\begin{abstract}
  We present the first next-to-next-to-leading logarithmic resummation
  for the two-jet rate in $e^+e^-$ annihilation in the Durham and
  Cambridge algorithms. The results are obtained by extending the {\tt
    ARES} method to observables involving any global, recursively
  infrared and collinear safe jet algorithm in $e^+e^-$ collisions. As
  opposed to other methods, this approach does not require a
  factorization theorem for the observables. We present predictions
  matched to next-to-next-to-leading order, and a comparison to LEP
  data.
\end{abstract}

\pacs{13.87.Ce,  13.87.Fh, 13.65.+i}

\maketitle

  
Jet rates and event shapes in electron-positron collisions played a
crucial role in establishing QCD as the theory of strong interactions,
see e.g.~\cite{Altarelli:1989ue,Bethke:1992gh}. Nowadays, these
observables are still among the most precise tools used for accurate
extractions of the main parameter of the theory, the strong coupling
constant $\alpha_s$. These fits rely on comparing precise measurements
of distributions to accurate perturbative predictions supplemented
with a modelling of non-perturbative effects.
Fixed order predictions up to next-to-next-to-leading order (NNLO) for
$e^+e^- \to$ 3 jets are
available~\cite{GehrmannDeRidder:2007hr,GehrmannDeRidder:2008ug,Weinzierl:2008iv,Weinzierl:2009ms,DelDuca:2016ily}.
However, they are not reliable in the two-jet limit, where the cross
section is dominated by multiple soft-collinear emissions. In this
region, terms as large as ${\cal O}(\alpha_s^n L^{2n})$ (where
$L=\ln(1/v)$) appear to all orders in the integrated distributions of
an observable $v$ that vanishes in the two-jet limit. These large
logarithms invalidate fixed-order expansions in the coupling constant
and reliable predictions can only be obtained by resumming the
logarithmically enhanced terms to all orders in $\alpha_s$. Double
logarithmic terms ${\cal O}(\alpha_s^n L^{2n})$ are known to
exponentiate (see e.g. ref.~\cite{Catani:1991hj}) and give rise to a
well-known Sudakov peak in differential distributions, where most of
the data lies.
For exponentiating observables, it is customary to define leading
logarithms (LL) as terms of the form $\alpha_s^n L^{n+1}$ for the
logarithm of the cross section, next-to-leading logarithms (NLL) as
$\alpha_s^n L^n$, next-to-next-to-leading logarithms (NNLL) as
$\alpha_s^n L^{n-1}$.
For several $e^+e^-$ observables, NNLL predictions (in some cases even
beyond) are nowadays
available~\cite{deFlorian:2004mp,Becher:2008cf,Chien:2010kc,Becher:2012qc,Hoang:2014wka,Banfi:2014sua,Becher:2015lmy,Frye:2016okc,Frye:2016aiz}.
On the contrary, two-jet rates have been described only at NLL
accuracy so far~\cite{Banfi:2001bz}.
The lack of precise theory predictions close to the peak of the
distribution limits the fit range that can be used to extract
$\alpha_s$ and results in larger perturbative uncertainties in the
latter.
Among the existing fits, extractions from the thrust and
$C$-parameter~\cite{Abbate:2010xh,Hoang:2015hka,Gehrmann:2012sc} that
rely on the most precise theory predictions show a tension with the world
average determination of the coupling~\cite{Bethke:2015etp}.
One of the issues is that at LEP energies non-perturbative corrections
are sizeable, and the separation between perturbative and
non-perturbative effects is subtle. Fits of $\alpha_s$ from the
two-jet rate have been so far performed based on pure
NNLO~\cite{Dissertori:2009qa}, or
NNLO+NLL~\cite{Bethke:2008hf,Dissertori:2009ik,OPAL:2011aa} results.
Owing to the different sensitivity to non-perturbative effects, an
extraction of $\alpha_s$ from NNLO+NNLL predictions for the two-jet
rate and from the vast amount of high-precision LEP
data~\cite{Heister:2003aj,Abdallah:2003xz,Adeva:1992gv,Achard:2004sv,Abbiendi:2004qz}
can shed light on this disturbing tension.
The aim of this letter is to present the first NNLL+NNLO results for
this observable.

The two-jet rate is defined through a clustering algorithm based on an
ordering $v_{ij}$ and a test variable $y_{ij}$. In the Durham
algorithm~\cite{Catani:1991hj} the two variables coincide
\begin{equation}
\label{eq:yijD}
y_{ij}^{(D)}=v_{ij}^{(D)}=2\frac{{\rm min}\{ E_i,E_j\}^2}{Q^2} \left( 1-\cos \theta_{ij}\right)\,,
\end{equation}
where $\theta_{ij}$ is the angle between (pseudo-)particles $i$ and
$j$, $E_i$ is the energy of the (pseudo-)particle $i$, and $Q$ is the
center-of-mass energy. The clustering procedure selects the pair with
the smallest $y_{ij}^{(D)}$. If the latter is smaller than a given
$\ycut$, the two particles are recombined into a pseudo-particle
according to some recombination scheme. Otherwise, the
clustering sequence stops, and the number of jets is defined as the
number of pseudo-particles left.
In the Cambridge algorithm~\cite{Dokshitzer:1997in,Bentvelsen:1998ug}, the test and
ordering variables differ, and are defined by
\begin{equation}
\label{eq:vijC}
y_{ij}^{(C)}=y_{ij}^{(D)}\,,\qquad 
v_{ij}^{(C)} = 2\left( 1-\cos \theta_{ij}\right)\,.
\end{equation}
The clustering procedure selects the pair with the smallest
$v_{ij}^{(C)}$. If the corresponding $y_{ij}^{(C)}$ is smaller than
$\ycut$, the two particles are recombined into a pseudo-particle,
otherwise the softer particle becomes a jet.  This is commonly
referred to as the soft freezing mechanism. The procedure stops when
no pseudo-particles are left.  The angular-ordered (AO) version of the
Durham algorithm~\cite{Dokshitzer:1997in} works identically to the
Cambridge algorithm, but without the freezing mechanism.
The three-jet resolution parameter $\ythree$ is defined as the minimum
$\ycut$ that produces two jets. The two-jet rate is the cumulative
integral of the $\ythree$ distribution, normalized to the total cross
section $\sigma$:
\begin{equation}
\label{eq:sigmacum}
\Sigma(\ycut)=\frac{1}{\sigma}\int_0^{\ycut} d \ythree \frac{d\sigma(\ythree)}{d\ythree}\,.
\end{equation}
The resummation technique formulated in ref.~\cite{Banfi:2014sua} for
event shapes does not require the factorization of the singular soft
and collinear modes in the observable's definition, but it rather
relies on a property known as recursive infrared and collinear (rIRC)
safety~\cite{Banfi:2004yd}. In this sense, the all-order treatment
does not require a factorization theorem for the
observable.\footnote{Note that a factorization theorem for the
  Cambridge algorithm is straightforward.}
In the following, we present an extension of the above method to jet
observables and apply it to the two-jet rate in the Durham and
Cambridge algorithm.

Let $\ythree(\{\tilde p\},k_1,\dots,k_n)$ denote a three-jet resolution which
depends on all $n+2$ final-state momenta, where $\{\tilde p\}$ indicates
the two Born momenta recoiling against the secondary emissions
$k_1,\dots,k_n$. Each parton $k_i$ is emitted off leg $\ell_i=1,2$.
The essence of the procedure described in ref.~\cite{Banfi:2014sua} is
that the NLL cross section is given by all-order configurations made
of partons independently emitted off the Born legs and widely
separated in angle~\cite{Banfi:2001bz}. The NNLL corrections are
obtained by correcting a {\it single} parton of the above ensemble to
account for all kinematic configurations that give rise to NNLL
effects~\cite{Banfi:2014sua}. 
The two-jet rate at NNLL can be written as 
\begin{align}
  \label{eq:y3-cs}
&  \Sigma(\ycut) = \,e^{-R_{\NNLL}(\ycut)} \left[
    \mathcal{F}_{\NLL}(\ycut) \right.\notag\\
&\left.+ \frac{\alpha_s(\mu_R)}{\pi}\delta{\mathcal F}_{\rm NNLL}(\ycut)
  \right],\notag\\
& \delta{\mathcal F}_{\rm NNLL}(\ycut) =\, \delta \mathcal{F}_{\rm clust} +
 \delta \mathcal{F}_{\rm correl} + \delta \mathcal{F}_{\rm sc}\notag\\&\hspace{2.1cm}+\delta
 \mathcal{F}_{\rm hc}
+\delta \mathcal{F}_{\rm rec} + \delta \mathcal{F}_{\rm wa} \,, 
\end{align}
where $\mu_R$ is the renormalization scale, and the physical origin of
the various contributions is discussed in the following.
The NNLL Sudakov radiator $R_{\NNLL}(\ycut)$ expresses the no-emission
probability above $\ycut$ and hence embodies the cancellation of
infrared and collinear divergences between the virtual corrections to
the Born process and the unresolved real emissions as defined in
ref.~\cite{Banfi:2014sua}. As such, it is inclusive over QCD radiation and
it is universal for all observables featuring the same scaling in the
presence of a single soft and collinear emission.
Since, in the soft-collinear limit,
$\ythree(\{\tilde p\},k) = (k_t/Q)^2$, where $k_t$ is the emission's
transverse momentum with respect to the emitting quark-antiquark pair,
one can obtain $R_{\NNLL}(\ycut)$ from appendix B of
ref.~\cite{Banfi:2014sua} by setting $a=2$ and taking the limit
$b_\ell\to 0$.
All remaining contributions in eq.~\eqref{eq:y3-cs} arise from
resolved real radiation in different kinematical regions.
In particular, the terms $\mathcal{F}_{\NLL}, \delta \mathcal{F}_{\rm
  sc}, \delta \mathcal{F}_{\rm clust},\delta \mathcal{F}_{\rm correl}$
originate from soft and collinear emissions. The function
$\mathcal{F}_{\NLL}$ is the only NLL correction to the radiator, and
it is defined in terms of soft and collinear gluons independently
emitted off the hard legs, and widely separated in rapidity. At NLL,
the upper rapidity bound is the same for all emissions and
approximated by $\ln (1/\sqrt{\ycut})$.  The soft-collinear term
$\delta \mathcal{F}_{\rm sc}$ arises from considering the NNLL effects
of the running coupling in the soft matrix element, as well as
restoring the exact rapidity bound for a single soft-collinear
emission.
The two functions $\delta \mathcal{F}_{\rm clust}$ and $\delta
\mathcal{F}_{\rm correl}$ account for configurations in which at most
two emissions are close in rapidity, and produce a pure abelian
clustering correction ($\delta \mathcal{F}_{\rm clust}$) and a
non-abelian correlated ($\delta \mathcal{F}_{\rm correl}$) one.
The hard-collinear ($\delta \mathcal{F}_{\rm hc}$) and recoil
($\delta \mathcal{F}_{\rm rec}$) corrections describe configurations
where one  emission of the ensemble is collinear, but hard. In
particular, $\delta \mathcal{F}_{\rm hc}$ takes into account the
correct approximation of matrix elements in this region, while
$\delta \mathcal{F}_{\rm rec}$ describes NNLL kinematical recoil
effects in the observable.
 Finally, the wide-angle correction $\delta \mathcal{F}_{\rm wa}$
 encodes configurations in which a single emission of the ensemble is
 soft and emitted at wide angles.

 All of the above corrections are obtained following a method close in
 spirit to an expansion by regions, i.e.\ by taking the proper
 kinematical limits in the squared amplitudes, the phase space and the
 observable constraint $\Theta\left(\ycut-\ythree(\{\tilde p
   \},k_1,\dots, k_n)\right)$.
 This leads to the definition of a tailored and simplified version of
 the observable - in our case a clustering algorithm - obtained from
 the exact one by taking the appropriate asymptotic limit in
 each kinematic region.
 The NNLL corrections that appear in eq.~\eqref{eq:y3-cs} have already
 been derived in the context of event-shapes
 resummations~\cite{Banfi:2014sua}, with the exception of the
 clustering correction $\delta \mathcal{F}_{\rm clust}$ which is
 absent for event-shapes, and the soft-collinear correction $\delta
 \mathcal{F}_{\rm sc}$ which is generalized in this letter. 
In the following we discuss the algorithms
 necessary to compute the NLL multiple emission function
 $\mathcal{F}_{\NLL}$ and the new correction $\delta \mathcal{F}_{\rm
   clust}$. The remaining algorithms are obtained following the same
 strategy of taking the asymptotic limit in the region considered in
 each correction. They are reported in ref.~\cite{BMMZadditional} both
 for the Durham and for the Cambridge.
 We will first discuss the case of the Durham algorithm, and we will
 eventually obtain the Cambridge result as a trivial case of the
 discussion that follows.\footnote{We note that the NNLL results
   presented in this letter are valid for all commonly used
   recombination schemes in $e^+e^-$ collisions (schemes $E$, $E_0$,
   $P$, $P_0$, cf.  ref.~\cite{Catani:1991hj} for their definition),
   while their NNLO counterpart depends on the recombination
   scheme.}

 We start by recalling the calculation of $\mathcal{F}_{\NLL}$, which
 is determined by an ensemble of soft-collinear, strongly angular-ordered
 partons emitted independently off the Born legs.  For soft emissions,
 recoil effects are negligible and all transverse momenta can be
 computed with respect to the emitting quark-antiquark pair. For each
 emission $k_i$ we define the rapidity fraction with respect to the
 emitting leg $\ell_i$ as
 $\xi_{i}^{(\ell_i)}=|\eta_i|/\ln(1/\sqrt{\ycut})$, where $\ln
 (1/\sqrt{\ycut})$ is the NLL rapidity bound, common to all emissions
 at this order. For this ensemble, the Durham algorithm is
 approximated by the following simplified version,
 $\yscbar{k_1,\dots,k_n}$~\cite{Banfi:2001bz}:
\begin{itemize}
\item[1.] Find the pseudo-particle $k_I$ with the smallest value of
  $\yscbar{k_I}=(k_{tI}/Q)^2$.
\item[2.] Considering only pseudo-particles $k_{j}$ collinear to the
  same leg $\ell$ as $k_I$, find the pseudo-particle $k_J$ which
  satisfies $\vec{k}_{tJ}\cdot\vec{k}_{tI}>0$ and has the smallest
  positive value of $\xi_{J}^{(\ell)}-\xi_I^{(\ell)}$.
\item[3.] If $k_J$ is found, recombine $k_I$ and $k_J$ into a new
  pseudo-particle $k_P$ with $\vec{k}_{tP}=\vec{k}_{tI}+\vec{k}_{tJ}$
  and $\xi_P^{(\ell)}=\xi_J^{(\ell)}$. Otherwise, $k_I$ is clustered
  with a Born leg, and removed from the list of pseudo-particles.
\item[4.] If only
  one pseudo-particle $k_P$ remains, then
  $\yscbar{k_1,\dots,k_n}=(k_{tP}/Q)^2$, otherwise go back to step 1.
\end{itemize}
Because of the assumption of strong rapidity ordering between the
emissions, this algorithm ensures that $\mathcal{F}_{\NLL}$ is free from
subleading effects.
We point out that as long as emissions are strongly ordered in rapidity, the
clustering history only depends on the rapidity ordering among
emissions, and not on the actual rapidities.  

The above algorithm is used whenever emissions are soft-collinear and
widely separated in angle, even beyond NLL order. In particular it can
be used to compute the NNLL soft-collinear correction $\delta
\mathcal{F}_{\rm sc}$.
This function is made of two contributions with different physical
origins:
\begin{equation}
\delta \mathcal{F}_{\rm sc} = \delta \mathcal{F}_{\rm sc}^{\rm rc} +
\delta \mathcal{F}_{\rm sc}^{\rm rap}.
\end{equation}
The term $\delta \mathcal{F}_{\rm sc}^{\rm rc}$ accounts for NNLL
effects in the coupling  which have been neglected in 
$\mathcal{F}_{\NLL}$, while the term
$\delta \mathcal{F}_{\rm sc}^{\rm rap}$ contains NNLL corrections due
to implementing the exact rapidity bound ($|\eta|<\ln(Q/k_{t})$) for a
{\it single} emission $k$ of the soft-collinear ensemble.
While the running-coupling correction
$\delta \mathcal{F}_{\rm sc}^{\rm rc}$ can be computed using the
strongly-ordered algorithm defined above, in complete analogy with
event-shape observables~\cite{Banfi:2014sua}, the rapidity correction
$\delta \mathcal{F}_{\rm sc}^{\rm rap}$ requires some care.
Since the exact rapidity bound for the emission $k$
($|\eta|<\ln(Q/k_{t})$) is larger than the NLL bound shared by the
other emissions $k_i$ ($|\eta_i|<\ln(1/\sqrt{\ycut})$), the rapidity
correction will be non-zero only if the rapidity of emission $k$ is,
in magnitude, the largest of all. The rapidity correction is then
computed by using the strongly-ordered algorithm defined above, with
emission $k$ fixed to be the most forward/backward of
all~\cite{BMMZadditional}.
Note that this issue is irrelevant for event shapes since they are
independent of the rapidity fractions, and that the derivation of the
rapidity correction given here can be equally applied in that case.

We now turn to the discussion of the NNLL clustering correction $\delta
\mathcal{F}_{\rm clust}$, which describes configurations in which at
most two of the independently-emitted, soft-collinear partons have
similar rapidities. We denote by $k_a$ and $k_b$ these two
emissions. The function $\delta\mathcal{F}_{\rm clust}$ accounts for the difference
between the observable $\ysc{k_a,k_b,k_1,\dots,k_n}$ in which $k_a$ and $k_b$
are close in rapidity, and the NLL observable $\yscbar{k_a,k_b,k_1,\dots,k_n}$
in which they are assumed to be far apart.
This correction appears whenever the observable depends on the
emissions' rapidity fractions, hence it is absent in the case of event
shapes.  Its formulation is analogous to the corresponding correction
derived for the jet-veto resummation in ref.~\cite{Banfi:2012jm}, and
is reported in~\cite{BMMZadditional}.

The algorithm that defines $\ysc{k_1,\dots,k_n}$ proceeds as the
NLL one, with an additional condition to be checked after
step 1:
\begin{itemize}
\item[1b.]  Let $k_{J_a}$ and $k_{J_b}$ be the pseudo-particles
  containing the partons $k_a$ and $k_b$. If these
  pseudo-particles are close in rapidity (i.e. if neither $k_a$ nor
  $k_b$ have been recombined with a pseudo-particle with larger
  $\xi^{(\ell)}$), check whether $k_{J_a}$ and $k_{J_b}$ cluster,
  i.e. if
\begin{equation}
\label{eq:yijDclust}
\min\{E_{J_a},E_{J_b}\}^2|\vec{\theta}_{J_a}-\vec{\theta}_{J_b}|^2 <\min\{k_{t J_a},k_{ t J_b}\}^2\,
\end{equation}
is satisfied, where $\vec\theta_i=\vec{k}_{ti}/E_i$.  If so, recombine
$k_{J_a}$ and $k_{J_b}$ by adding transverse momenta vectorially, and setting
the rapidity fraction of the resulting pseudo-particle $k_J$ to $\xi_J^{(\ell)}\simeq
\xi_{J_a}^{(\ell)}\simeq \xi_{J_b}^{(\ell)}$.
\end{itemize}
The same algorithm is employed in the computation of the NNLL
correlated correction
$\delta \mathcal{F}_{\rm correl}$~\cite{Banfi:2014sua}
(see~\cite{BMMZadditional} for details).
 \begin{figure*}[htp]
  \centering
  \includegraphics[width=0.47\linewidth]{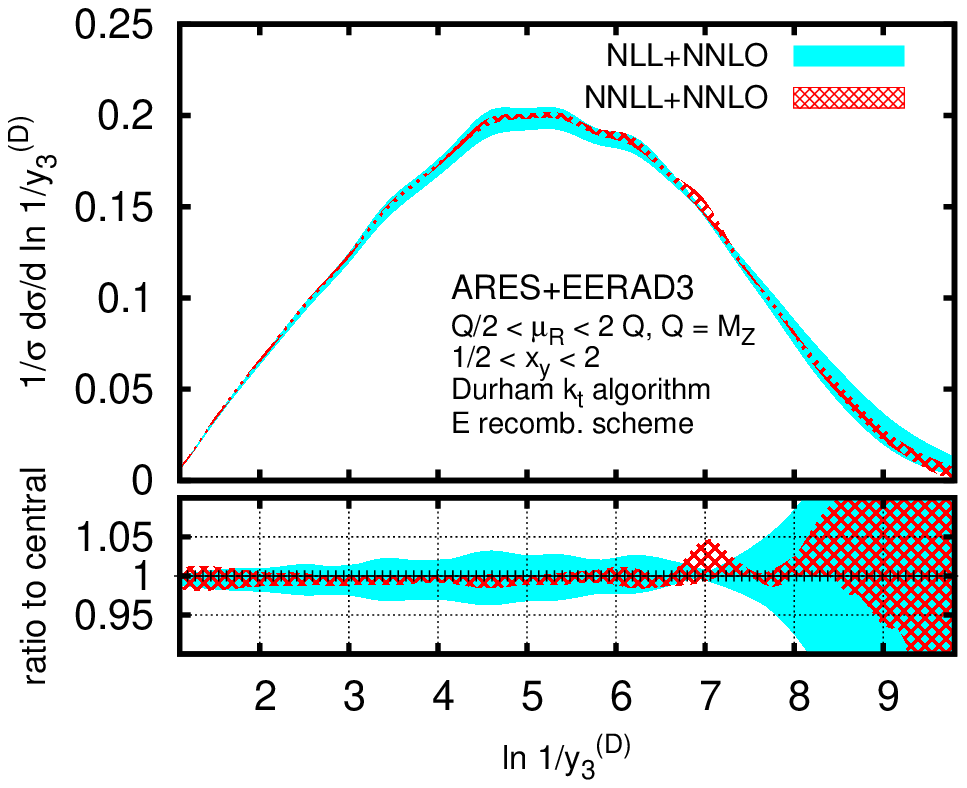}
  \includegraphics[width=0.47\linewidth]{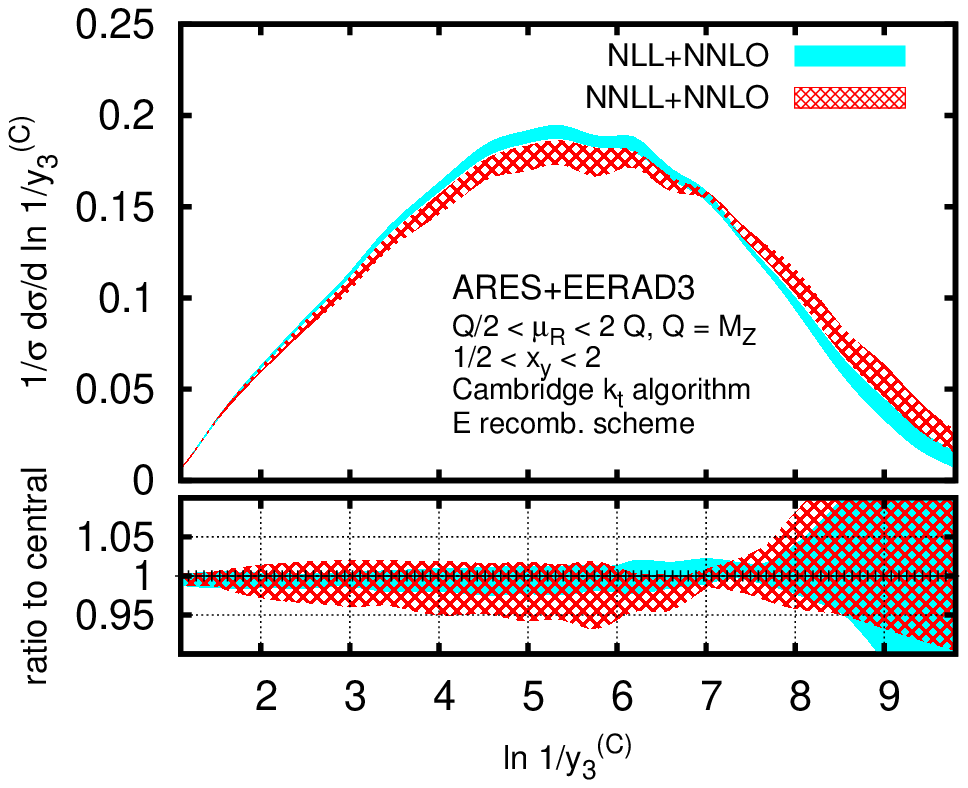}
\caption{%
  Differential distributions for the three-jet resolution in the
  Durham (left) and Cambridge (right) algorithms. The plots show both
  the NLL+NNLO (blue/solid) and the NNLL+NNLO (red/hatched) results.}
  \label{fig:matched}
\end{figure*}
In a similar way we approximate the original algorithm to compute the
remaining NNLL corrections whose definition follows exactly the one
given for event shapes~\cite{Banfi:2014sua}.

The considerations made so far for the Durham case can be
straightforwardly adapted to any other rIRC jet algorithm.
In particular, for the Cambridge algorithm the NNLL logarithmic
structure is much simpler. In this case the ordering
variable~\eqref{eq:vijC} only depends on the angular distance between
emissions. Since at NLL all partons are well separated in rapidity,
there will be no clustering between the emissions, and each of them
will be recombined with one of the Born legs in an angular-ordered
way. One therefore obtains the trivial result
${\cal F}_{\rm NLL}(\lambda)=1$. The same arguments imply that the
NNLL corrections
$\delta \mathcal{F}_{\rm sc} =\delta \mathcal{F}_{\rm
  hc}=0$~\cite{BMMZadditional}.
Moreover, both the recoil and the wide-angle corrections admit a
simple analytic form given that the emission emitted either at wide
angles or collinearly will never cluster with any of the other
soft-collinear emissions. As a consequence the
contribution from this emission factorizes with respect to the
remaining ensemble~\cite{BMMZadditional}.
The same property applies to
the clustering and correlated corrections which can be entirely
formulated in terms of the clustering condition between two
soft-collinear emissions~\cite{BMMZadditional}, analogously to the jet veto
resummation~\cite{Banfi:2012jm}.
We note that the freezing condition present in the Cambridge algorithm
does not play a role at NNLL. Therefore the AO version of the Durham
algorithm coincides with the Cambridge algorithm at this order, while
the two differ at NNLO.

We tested our results by subtracting the derivative of the
second-order expansion of eq.~\eqref{eq:y3-cs} from the
${\cal O}(\alpha_s^2)$ distributions obtained with the generator {\tt
  Event2}~\cite{Catani:1996vz}, finding
agreement~\cite{BMMZadditional}. Moreover, we applied the method to
both the inclusive-$k_t$~\cite{Weinzierl:2010cw,Cacciari:2011ma} and
the flavor-$k_t$~\cite{Banfi:2006hf} algorithms, finding also perfect
agreement with {\tt Event2} at ${\cal O}(\alpha_s^2)$.\footnote{A
  check at ${\cal O}(\alpha_s^3)$ would require a very stable
  fixed-order distribution at small $y_{\rm cut}$ at this
  order. However, we have not been able to obtain stable enough
  predictions to carry out this test.}

\begin{figure}[b]
  \centering
  \includegraphics[width=0.9\linewidth]{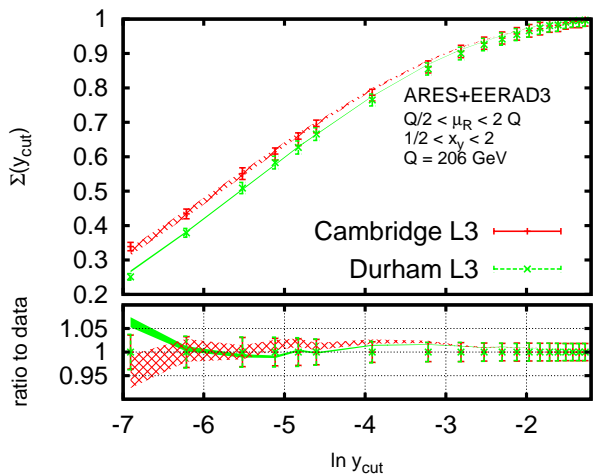}
 \caption{Comparison of NNLL+NNLO predictions for the two-jet rates
   to data from the L3 collaboration~\cite{Achard:2004sv}.}
  \label{fig:NNLL-L3}
\end{figure}

We illustrate the impact of our calculation by matching the NNLL
two-jet rate~\eqref{eq:y3-cs} to the ${\cal O}(\alpha_s^3)$
result obtained with the program {\tt EERAD3}~\cite{Ridder:2014wza}
for both the Durham and the Cambridge
algorithms. Figure~\ref{fig:matched} shows the matched differential
distributions for the three-jet resolution parameter, defined
in~\eqref{eq:sigmacum}, at NNLL+NNLO and NLL+NNLO. The results are
obtained at $Q=M_Z$, using the coupling $\alpha_s(M_Z)=0.118$, and the
$E$ recombination scheme.
To impose unitarity, following ref.~\cite{Banfi:2014sua}, we employ
the modified logarithms
\begin{equation}
\ln\frac{1}{y_3} \to 
\ln
\left(1
+\left(\frac{x_y}{y_3}\right)-
\left(\frac{x_y}{y_{\rm 3, max}}\right)
\right)\,, 
\end{equation}
in such a way that the $x_y$ dependence is N$^3$LL. This also ensures
that the distribution vanishes at the kinematical endpoint
$y_{\rm 3, max}$, taken from the NNLO result. Furthermore, the
variation of $x_y$ probes the size of subleading logarithmic effects.
Our theoretical uncertainties are obtained by varying, one at the
time, $x_y$ and the renormalization scale $\mu_R$ by a factor of two
in either direction around the central values $x_y=1$ and $\mu_R = Q$,
and taking the envelope of these variations.

For the Durham algorithm, as expected, we observe a significant
reduction of the theory error when going from NLL to NNLL. On the
contrary, for the Cambridge algorithm, NNLL corrections are quite
large, and the NNLL uncertainty is larger than the NLL one, which in
turn seems to be underestimated. This effect can be understood by
observing that the NLL prediction for the Cambridge algorithm does not
contain any information about multiple emissions effects since no
clustering occurs at this order and ${\cal F}_{\rm NLL}=1$. These
effects appear only at NNLL, explaining the sizable numerical
corrections. It follows that the NLL theory uncertainty as estimated
in figure~\ref{fig:matched} is unable to capture large subleading
effects. A similar phenomenon was already observed in the resummation
for the jet-veto efficiency~\cite{Banfi:2012jm}.

To conclude, in figure~\ref{fig:NNLL-L3} we compare our NNLL+NNLO
prediction to the data taken by the L3 collaboration at
LEP2~\cite{Achard:2004sv} at $Q=206$\,GeV.  At this high
center-of-mass energy the impact of hadronization effects, which are
not included in our calculation, is moderate. Overall, we find good
agreement with data down to the lowest values of $\ycut$. Owing to the
small residual perturbative uncertainties, our calculation shows
promise for a precise determination of the strong coupling using
$e^+e^-$ data measured at LEP.

In this paper we have presented a general method for final-state
resummation at NNLL order for global rIRC safe observables that vanish
in the two-jet limit, where a single family of large logarithms is
resummed. We derived explicit results for the two-jet rate in
$e^+e^-$. The computer code {\tt ARES} used to obtain the results
presented here can be made available upon request to the authors.

\begin{acknowledgments} 
We would like to thank Gavin Salam for fruitful discussions. 
PM and GZ have been partially supported by the ERC grant 614577 {\it
  HICCUP}.
The work of PM is partly supported by the SNF under grant
PBZHP2-147297, and the work of AB is supported by the STFC under grant
number ST/L000504/1.
We gratefully acknowledge the Mainz Institute for Theoretical Physics
(MITP) (PM and GZ), KITP (GZ), and the CERN's Theory Department (AB,
HM, PM) for hospitality and partial support while part of this work
was carried out. AB, HM and PM acknowledge the use of the DiRAC
Complexity HPC facility under the grant PPSP62.
\end{acknowledgments} 


\newpage

\onecolumngrid
\newpage
\appendix

\section*{Supplemental material}

\makeatletter
\renewcommand\@biblabel[1]{[#1S]}
\makeatother

\setcounter{figure}{0}
We provide here explicit formulae that complete the discussion of the
letter. Furthermore analytic results for the case of the
Cambridge algorithm are derived explicitly.

\subsection{Next-to-next-to-leading-logarithmic real corrections}

\subsubsection{Resolved real corrections at NLL}
\label{sec:NLL-recap}

At NLL accuracy the details of the resolved real radiation are
described by the multiple emission function $\FNLL$.  $\FNLL$ is
defined on an ensemble of independently-emitted soft and collinear
partons, widely separated in rapidity. Moreover, all emissions have
the same rapidity bound $|\eta_i| < \ln(1/\sqrt{\ycut})$. The multiple
emission function depends on $\lambda = \alpha_s(\mu_R) \beta_0
\ln(1/\ycut)$ ($\mu_R$ being the renormalization scale) with $\beta_0
= \frac{11C_A - 2 n_f}{12\pi}$, and is defined as
\begin{equation}
  \label{eq:F-nll}
  \FNLL(\lambda) = \int \dZ \, \Theta\left(1-\lim_{\ycut\to 0}\frac{
\yscbar{\{k_i\}}
}{\ycut}\right)\,.
\end{equation}
In the above equation, $\dZ$ is the soft-collinear measure, which is
defined for any arbitrary function $G(\{\tilde p\},k_1,\dots,k_n)$ as 
\begin{equation}
\label{eq:dZ}
\begin{split}
\int \dZ  G(\{\tilde p\},\{k_i\})=\epsilon^{R'_{\mathrm{NLL}}}
   \sum_{n=0}^{\infty}\frac{1}{n!} \prod_{i=1}^n
    \int_{\epsilon}^{\infty} \frac{d\zeta_i}{\zeta_i}\int_0^{2\pi}
   \frac{d\phi_i}{2\pi} \sum_{\ell_i=1,2} \int_0^1 d\xi_i^{(\ell_{i})}\,R'_{\mathrm{NLL}, \ell_i}G(\{\tilde p\},k_1,\dots,k_n)\,.
\end{split}
\end{equation}
Here $\zeta_i = \frac{k_{ti}^2}{Q^2 \ycut}$, $\xi_i^{(\ell_i)}=|\eta_i|/\eta_{\max}$, 
and $R'_{\mathrm{NLL}, \ell}$ is  defined in appendix B of
ref.~\cite{Banfi:2014sua}. 
For each emission $k_i$ the sum is over the two emitting legs $\ell_i =
1,2$, and $\ell_i = 1$ ($\ell_i = 2$) when $\eta_i$ is positive
(negative).  
The measure~\eqref{eq:dZ} differs from that used in the case of event
shapes, because of the presence of the integrals over the rapidity
fractions $\xi_i^{(\ell_i)}$, where in this case
$\eta_{\max}=\ln(1/\sqrt{y_{\rm cut}})$.  In the case of event-shapes,
one could integrate inclusively over the rapidity fractions. This is
not the case for $\ythree$ since it depends explicitly on the
particles' $\xi_i^{(\ell_i)}$. $\dZ$ satisfies the normalization
condition
\begin{equation}
  \label{eq:dZ-norm}
  \int \dZ \prod_i  \Theta\left(1-\lim_{\ycut\to 0}\frac{
\yscbar{k_i}}{\ycut}\right)=1\,. 
\end{equation}
Note that in the presence of a single soft-collinear emission 
$\yscbar{k_i} = \ysc{k_i}$. 
\begin{itemize}
\item {\bf Durham algorithm:} when emissions are both soft and
  collinear, and strongly ordered in rapidity, the observable, denoted
  by $\yscbar{k_1,\dots,k_n}$, can be computed with the following
  simplified algorithm:
\begin{itemize}
\item[1.] Find the index $I$ of the smallest
  $\yscbar{k_I}=\ysc{k_I}=(k_{tI}/Q)^2$.
\item[2.] Considering only pseudo-particles $k_{j}$ collinear to the
  same leg $\ell$ as $I$, find parton $k_J$ which satisfies
  $\vec{k}_{tJ}\cdot\vec{k}_{tI}>0$ and has the smallest {\em positive} value of
  $\xi_{J}^{(\ell)}-\xi_I^{(\ell)}$.
\item[3.] If $k_J$ is found, recombine partons $I$ and $J$ into a new
  pseudo-particle $k_P$ with $\vec{k}_{tP}=\vec{k}_{tI}+\vec{k}_{tJ}$
  and $\xi_P^{(\ell)}=\xi_J^{(\ell)}$. 
  Otherwise, $k_I$ is clustered with a Born leg, and removed from the list of
  pseudo-particles.
\item[4.] If only
  one pseudo-particle $k_P$ remains, then
  $\yscbar{k_1,\dots,k_n}=(k_{tP}/Q)^2$, otherwise go back to step 1.
\end{itemize}

\item {\bf Cambridge algorithm:} since for the Cambridge algorithm no
  recombinations occur if all emissions are widely separated in
  rapidity, we have:
\begin{equation}
  \label{eq:y3sc-Cambridge}
 \Theta\left(1-\lim_{\ycut\to 0} \frac{\yscbar{k_1,\dots,k_n}}{\ycut}\right)=\prod_{i=1}^n  \Theta\left(1-\lim_{\ycut\to 0}\frac{
\ysc{k_i}}{\ycut}\right)\,.
\end{equation}
Therefore, for this algorithm, $\FNLL(\lambda)=1$.
\end{itemize}

\subsubsection{Soft-collinear correction}
The soft-collinear NNLL correction takes into account the correct
rapidity bound for one of the soft-collinear emissions that give rise
to the NLL multiple emission function, as well as NNLL contributions
arising from the running of the QCD coupling in the soft-collinear
matrix elements. We denote by $k$ the emission for which we account
for either effect, and introduce $\zeta$ such that
$\ysc{k}=k_t^2/Q^2=\zeta \ycut$. If $\ythree$ were an event shape, we could integrate inclusively over the rapidity fraction of each emission. As a result, the emission probability for $k$, collinear to the Born leg $\ell$, would be proportional to the function $R'_\ell\left(\zeta \ycut\right)$ defined in Section 2 of ref.~\cite{Banfi:2014sua}. 
In this case both NNLL effects could be accounted for by expanding
$R'_\ell\left(\zeta y_{\rm cut}\right)$ as follows:
\begin{equation}
\label{eq:rpexp}
  R'_\ell\left(\zeta y_{\rm cut}\right) \simeq R'_{\rm NLL,\ell}(y_{\rm cut})+\delta R'_{\rm NNLL,\ell}(y_{\rm cut})+R''_{\ell}(y_{\rm cut}) \ln\frac{1}{\zeta}\,.
\end{equation} The full expressions for
$\delta R'_{\rm NNLL,\ell}(y_{\rm cut})$ and $R''_{\ell}(y_{\rm cut})$
are given in ref.~\cite{Banfi:2014sua}. 
In the present case, this correction must be formulated in a slightly
more general way than the corresponding one defined for event-shape
observables~\cite{Banfi:2014sua}.
The NNLL term proportional to $\delta R'_{\rm NNLL,\ell}(y_{\rm cut})$ in eq.~\eqref{eq:rpexp} 
contains the contribution from the one-loop cusp anomalous dimension as well as from the two-loop running of the QCD coupling. In this term, the rapidity of all emissions is bounded by the NLL limit $\ln(1/\sqrt{\ycut})$. Therefore this correction is unchanged with respect to event shapes, and gives rise to 
\begin{equation}
  \label{eq:dF-sc}
  \begin{split}
&\frac{\pi}{\alpha_s(\mu_R)}
\int_0^\infty
    \frac{d\zeta}{\zeta}  
    \int_0^{2\pi}\frac{d\phi}{2\pi} \sum_{\ell=1,2} \int_0^1 d\xi^{(\ell)} \delta R'_{{\rm NNLL},\ell}\int \dZ \times \\ & \times 
\left[\Theta\left(1-\lim_{\ycut\to
        0}\frac{\yscbar{k, \{k_i\}}}{\ycut}\right)-\Theta(1-\zeta)\Theta\left(1-\lim_{\ycut\to
        0}\frac{\yscbar{\{k_i\}}}{\ycut}\right)\right]\,,
  \end{split}
\end{equation}
where the soft-collinear observable $\yscbar{k, \{k_i\}}$ is computed
by means of the NLL algorithms given in the previous section.

The remaining term in the r.h.s. of eq.~\eqref{eq:rpexp} is proportional to the function $R''_{\ell}(y_{\rm cut})$ given by
\begin{equation}
\label{eq:rs}
R''_{\ell}(y_{\rm cut}) = \frac{\alpha_s(\sqrt{\ycut} Q)}{2 \pi}C_F\left( \beta_0 \alpha_s(\sqrt{\ycut} Q)
\ln\left(\frac{1}{y_{\rm cut}}\right) + 1\right)\,. 
\end{equation}
The above function is made of two contributions: the term proportional to $\beta_0$ arises from expanding $\alpha_s(k_t)$ around $\alpha_s(\sqrt{\ycut}Q)$ in the soft emission matrix element as follows
\begin{align}
  \label{eq:asexp}
  \alpha_s(k_{t})\simeq \alpha_s(\sqrt{\ycut} Q) + \beta_0\alpha_s^2(\sqrt{\ycut} Q)\ln\frac{1}{\zeta}\,.
\end{align}
The term proportional to $\ln(1/\zeta)$ is purely NNLL. Therefore,
when integrating over the emissions' phase space, we can set all
rapidity bounds to the NLL limit $\ln (1/\sqrt{\ycut})$, neglecting
subleading logarithmic terms. This approximation is identical to the
one defining eq.~\eqref{eq:dF-sc}, therefore the two corrections can
be put together to define the running-coupling part
$\delta\mathcal{F}_{\rm sc}^{\rm \,rc}$ of the soft-collinear
correction as follows:
\begin{equation}
  \label{eq:dF-rc}
  \begin{split}
\delta\mathcal{F}_{\rm sc}^{\rm \, rc}(\lambda) & = 
\frac{\pi}{\alpha_s(\mu_R)}
\int_0^\infty
    \frac{d\zeta}{\zeta}  
    \int_0^{2\pi}\frac{d\phi}{2\pi} \sum_{\ell=1,2} \int_0^1 d\xi^{(\ell)} \left(\delta R'_{{\rm NNLL},\ell} + \lambda R''_{\ell} \ln\frac{1}{\zeta}\right)\int \dZ \times \\ & \times 
\left[\Theta\left(1-\lim_{\ycut\to
        0}\frac{\yscbar{k, \{k_i\}}}{\ycut}\right)-\Theta(1-\zeta)\Theta\left(1-\lim_{\ycut\to
        0}\frac{\yscbar{\{k_i\}}}{\ycut}\right)\right]\,,
  \end{split}
\end{equation}
where 
\begin{equation}
\lambda R''_{\ell} = \frac{C_F}{2\pi} \beta_0
\alpha_s^{2}(\sqrt{\ycut} Q) \ln\frac{1}{y_{\rm cut}}\,.
\end{equation}

The second term in eq.~\eqref{eq:rs} is associated with the correct
rapidity bound for emission $k$.
Given that the observable in this case depends on the rapidity
fractions of the emissions, unlike for event shapes the latter
correction is not accounted for by eq.~\eqref{eq:rpexp}.

To study how the form of this correction is modified, let us consider
a given ensemble of $n$ emissions $k_1,\dots,k_n$ strongly ordered in
rapidity, collinear to the same hard leg, say $\ell=1$ which
corresponds to positive rapidities. All of the emissions have the NLL
rapidity bound $\ln (1/\sqrt{\ycut})$ except for the emission $k_j$
which has the exact rapidity bound
$\ln(Q/k_{tj}) > \ln(1/\sqrt{\ycut})$. The latter relation can be
proven by observing that for all emissions $k_i$ one has that
$k_{ti}\leq \sqrt{\ycut}Q$. This statement is trivial if no clustering
occurs. If pseudo-particles $k_I$ and $k_J$ are recombined, the
transverse momentum of the resulting jet
$|\vec{k}_{tI} + \vec{k}_{tJ}|$ will be larger than $k_{tI}$ and
$k_{tJ}$. This is because a clustering occurs only if
$\vec{k}_{tI}\cdot \vec{k}_{tJ} > 0$ in the NLL algorithm. By
induction, in all configurations which end up with two jets
(i.e. $\yscbar{\{k_i\}} < \ycut$), one has $k_{ti}\leq \sqrt{\ycut}Q$
for all particles $k_i$.

Let us consider a given ordering of transverse momenta $\{k_{ti}\}$ of
the $n$ emissions. For such a configuration of transverse momenta,
$n!$ rapidity orderings are available. Each rapidity ordering
corresponds to a different value for the observable in its NLL version
(see the algorithm given in the previous section).  We now assume that
all emissions but $k_j$ have the NLL rapidity bound
$\ln (1/\sqrt{\ycut})$, whereas $\eta_j < \ln (Q/k_{tj})$.
Without loss of generality, we start by considering the generic
ordering $\eta_1 > \eta_2 >\dots > \eta_j>\dots> \eta_n$.
We can identify two possible scenarios: when the most forward emission
has rapidity $\eta_1 < \ln (1/\sqrt{\ycut})$, and when
$\ln (1/\sqrt{\ycut}) < \eta_1 < \ln (Q/k_{t1}) $.
In the first case, after including running couplings and color
factors, the corresponding rapidity integral is
\begin{align}
\label{eq:rap-integral-1}
  I_1^{(n)}=&\left(\frac{C_F}{\pi}\right)^n\prod_{i=1}^{n}\alpha_s(k_{ti})\int^{\ln (1/\sqrt{\ycut})} d\eta_1 \int^{\eta_1}
  d\eta_2\dots \int^{\eta_{j-1}} \!\! d\eta_j\dots \int^{\eta_{n-1}} \!\! d\eta_n= \left(\frac{C_F}{\pi}\right)^n\prod_{i=1}^{n}\alpha_s(k_{ti})                    \frac{1}{n!}\ln^{n}\frac{1}{\sqrt{y_{\rm cut}}}\,.
\end{align}
We stress that this result is the same regardless of the rapidity
bound of emissions $k_2,\cdots,k_n$. Note that the integral in
eq.~\eqref{eq:rap-integral-1} is correct under the assumption of
strong rapidity ordering. The extra NNLL correction originating from
configurations in which two emissions are close in rapidity, for which
the NLL version of the observable cannot be applied, is taken into
account in the clustering corrections derived below.
It is manifest that the integral~\eqref{eq:rap-integral-1} contributes
to a given kinematic configuration starting at NLL. To neglect
subleading effects, we can expand the strong coupling in
eq.~\eqref{eq:rap-integral-1} as in eq.~\eqref{eq:asexp}. This
leads to
\begin{align}
\label{eq:soft-collinear-form-1}
  I_1^{(n)} &= \left(\frac{C_F}{\pi}\right)^n\alpha_s^{n}(\sqrt{\ycut}Q) \frac{1}{n!}\ln^{n}\frac{1}{\sqrt{y_{\rm
  cut}}} + \beta_0 \alpha_s^{n+1}(\sqrt{\ycut} Q) \left(\frac{C_F}{\pi}\right)^n\frac{1}{n!}\ln^{n}\frac{1}{\sqrt{y_{\rm
  cut}}} \sum_{i=1}^{n}\ln\frac{1}{\zeta_i}  + {\cal O}({\rm N^3LL})\notag\\
&\simeq\frac{(R'_{\ell}(\ycut))^n}{n!} + \lambda R''_{\ell}(\ycut)\frac{(R'_{\ell}(\ycut))^{n-1}}{n!}\sum_{i=1}^{n}\ln\frac{1}{\zeta_i}\,,
\end{align}
where we used
\begin{equation}
  \ln\frac{Q}{k_{ti}}=\ln\frac{1}{\sqrt{\ycut}} +
  \ln\frac{1}{\sqrt{\zeta_i}}\,,
\end{equation}
and $\zeta_i = (k_{ti}/Q)^2/y_{\rm cut}$.

Analogously, the configurations in
which $\ln (1/\sqrt{\ycut}) < \eta_1 < \ln (Q/k_{t1}) $ lead to
\begin{align}
\label{eq:rap-integral-2}
 I_{2}^{(n)}=\left(\frac{C_F}{\pi}\right)^n\prod_{i=1}^{n}\alpha_s(k_{ti})\int^{\ln (Q/k_{t1})}_{\ln (1/\sqrt{\ycut})} d\eta_1 \int^{\eta_1}
  d\eta_2\dots \int^{\eta_{j-1}}\!\! d\eta_j\dots \int^{\eta_{n-1}}\!\! d\eta_n\,.
\end{align}
The bound in $\eta_2$ can be replaced with $\ln (1/\sqrt{\ycut})$
since the region where $\eta_2 > \ln (1/\sqrt{\ycut})$ gives rise to a
subleading correction. Moreover, the argument of the running coupling
can be replaced with $\sqrt{\ycut} Q$ for all emissions at NNLL. With
these replacements we have
\begin{align}
\label{eq:rap-integral-3}
 I_{2}^{(n)}=\left(\frac{C_F}{\pi}\right)^n\alpha_s^{n}(Q\sqrt{\ycut})
                 \frac{1}{(n-1)!}\ln^{n-1}\frac{1}{\sqrt{\ycut}}\ln\frac{1}{\sqrt{\zeta_1}} = (1-\lambda) R''_{\ell}(\ycut)\frac{(R'_{\ell}(\ycut))^{n-1}}{(n-1)!}\ln\frac{1}{\zeta_1}\,.
\end{align}
Eq.~\eqref{eq:rap-integral-3} gives a pure NNLL contribution, and it
is obtained in the limit of strong rapidity ordering. The
configuration in which two emissions are close in rapidity here gives
a subleading correction, proving that there is no overlap with the
configurations contributing to the clustering correction.

In eq.~\eqref{eq:soft-collinear-form-1} we can recognise the NLL
contribution (first term in the r.h.s.) that gives rise to the
function ${\cal F}_{\rm NLL}$, and the NNLL correction proportional to
$\lambda R''_{\ell}$ in eq.~\eqref{eq:dF-rc}, that starts at
${\cal O}(\alpha_s^3)$. Eq.~\eqref{eq:rap-integral-3} gives rise to a
pure NNLL correction which accounts for the exact rapidity bound for a
single emission. At NNLL accuracy, this bound matters only for the
most forward/backward emission. We denote this correction by
$\delta\mathcal{F}_{\rm sc}^{\rm \, rap}$.

In order to compute the latter to all orders, we set emission $k$, the
one with the correct bound, to be the most forward/backward, and we
generate randomly the rapidity fractions of the remaining
emissions. This gives the following correction
\begin{equation}
  \label{eq:dF-sc3}
  \begin{split}
    \delta\mathcal{F}_{\rm sc}^{\rm\, rap}(\lambda) & =
    \frac{\pi}{\alpha_s(\mu_R)} \int_0^\infty \frac{d\zeta}{\zeta}
    \int_0^{2\pi}\frac{d\phi}{2\pi} \sum_{\ell=1,2}
    (1-\lambda)R'_{\ell}\ln\frac{1}{\zeta}\int
    \dZ \times \\ & \times \left[\Theta\left(1-\lim_{\ycut\to
          0}\frac{\yscbar{k,
            \{k_i\}}}{\ycut}\right)-\Theta(1-\zeta)\Theta\left(1-\lim_{\ycut\to
          0}\frac{\yscbar{\{k_i\}}}{\ycut}\right)\right]_{\xi^{(\ell)}
      = 1}\,,
  \end{split}
\end{equation}
where now $\zeta,\xi^{(\ell)},\phi$ refer to the emission $k$ with
exact rapidity bound, and
$(1-\lambda)R''_{\ell} = C_F \alpha_s(\sqrt{\ycut}
Q)/(2\pi)$~\cite{Banfi:2014sua}.
The condition $\xi^{(\ell)} = 1$ indicates the rapidity fraction of
$k$ has been fixed to $1$ reflecting the fact that the emission with
the correct rapidity bound must be the most forward/backward in
rapidity.

In the case of the event shapes, the integrals over the rapidity
fractions can be evaluated inclusively, and the sum
\begin{equation}
\delta\mathcal{F}_{\rm sc}^{\rm\, rc}(\lambda) + \delta\mathcal{F}_{\rm sc}^{\rm\, rap}(\lambda)
\end{equation}
reproduces the soft-collinear correction formulated in
ref.~\cite{Banfi:2014sua}. Therefore, the formulation given here can
be easily adapted to other observables, including event shapes.

\begin{itemize}
\item {\bf Cambridge algorithm:} the form of the soft-collinear
  corrections $\delta\mathcal{F}_{\rm sc}$ can be simplified using
  eq.~\eqref{eq:y3sc-Cambridge} as 
\begin{align}
  \label{eq:dF-sc-cam}
&\Theta\left(1-\lim_{\ycut\to
        0}\frac{\yscbar{k, \{k_i\}}}{\ycut}\right)-\Theta(1-\zeta)\Theta\left(1-\lim_{\ycut\to
        0}\frac{\yscbar{\{k_i\}}}{\ycut}\right)\notag\\ &=
   \left[ \Theta\left(1-\lim_{\ycut\to 0} \frac{\yscbar{k}}{\ycut}\right)  -
    \Theta(1-\zeta)\right] \prod_{i=1}^n  \Theta\left(1-\lim_{\ycut\to 0}\frac{
\ysc{k_i}}{\ycut}\right)= 0\,,
\end{align}
where we made use of the definition of
$\zeta=\ysc{\{k\}}=\zeta \ycut$. This result trivially leads to
$\delta\mathcal{F}_{\rm sc}(\lambda)=0$ for the Cambridge algorithm.
\end{itemize}

\subsubsection{Clustering corrections}
\label{sec:nnll-clust}
This correction describes an ensemble of soft-collinear partons
emitted off the Born legs of which at most two are close in rapidity
and the remaining ones are strongly separated in angle. For rIRC safe
observables, this kinematical configuration contributes only at NNLL
order and beyond. The clustering correction
$\delta\mathcal{F}_{\rm clust}$ encodes the abelian contribution to
the above configuration, where the two partons which are close in
rapidity have been emitted independently. Its expression reads
\begin{equation}
  \label{eq:F1-clustering}
  \begin{split}
    \delta\mathcal{F}_{\rm clust}(\lambda)&=
    \frac{1}{2!}\int_0^{\infty}\frac{d\zeta_a}{\zeta_a}\int_{0}^{2\pi}\frac{d\phi_a}{2\pi}\sum_{\ell_a=1,2}\int_0^1
    d\xi_a^{(\ell_a)}\left(\frac{2 C_F\lambda}{\beta_0}\frac{R^{''}_{\ell_a}(\ycut)}{\alpha_s(\mu_R)}\right)
    \int_0^{\infty} \frac{d\kappa}{\kappa}
    \int_{-\infty}^{\infty}\!\!\! d\eta
    \int_{0}^{2\pi}\frac{d\phi}{2\pi} \times\\& \times \int \dZ
    \left[\Theta\left(1-\lim_{\ycut\to
          0}\frac{\ysc{k_a,k_b,\{k_i\}}}{\ycut}\right)-\Theta\left(1-\lim_{\ycut\to
          0}\frac{\yscbar{k_a,k_b,\{k_i\}}}{\ycut}\right)\right]\,,
  \end{split}
\end{equation}
where $\ysc{k_a,k_b,\{k_i\}}$ is obtained with the clustering
procedures outlined below for this type of kinematical configuration,
while $\yscbar{k_a,k_b,\{k_i\}}$ is the NLL version of the
algorithm. We have parameterized the phase space of the emission $k_b$
in terms of the variables
\begin{equation}
\label{eq:corvars}
\kappa=k_{t,b}/k_{t,a}\,\qquad \eta=\eta_b-\eta_a\,, \qquad 
\phi=\phi_b-\phi_a\,.
\end{equation}
In terms of these variables $k_b$ can be written as 
\begin{equation}
  \label{eq:kbarb}
   k_b = \kappa\, Q \, \sqrt{\zeta_a \ycut}
(\cosh(\eta_a+\eta),\cos(\phi_a+\phi),\sin(\phi_a+\phi),\sinh(\eta_a+\eta))\,.
\end{equation}
In order to eliminate subleading effects, in the calculation of the
observable we impose that $k_b$ belongs to the same hemisphere as
$k_a$. In practice, this is accomplished by setting $\ell_b=\ell_a$
and
$\xi_b^{(\ell_a)}= \xi_a^{(\ell_a)}+\mathrm{sign}(\eta) \delta\xi$,
with $\delta \xi$ an arbitrarily small quantity. Unlike for the jet
rates, event shapes are independent of the rapidity fractions of the
emissions, therefore this correction is absent for such observables.
\begin{itemize}
\item {\bf Durham algorithm:}  the resulting algorithm goes along the
  lines of the strongly-ordered one, with an additional condition to
  be checked after step 1.
  \begin{itemize}
  \item[1b.]  Let $k_{J_a}$ and $k_{J_b}$ be the pseudo-particles
    containing the partons $k_a$ and $k_b$. If the latter
    pseudo-particles are close in rapidity (i.e.~if neither $k_a$ nor
    $k_b$ have been recombined with a pseudo-particle with larger
    $\xi^{(\ell)}$), check whether $k_{J_a}$ and $k_{J_b}$ cluster, i.e.~if
    \begin{equation}
      \label{eq:yijDclust-app}
      \min\{E_{J_a},E_{J_b}\}^2|\vec{\theta}_{J_a}-\vec{\theta}_{J_b}|^2 <\min\{k_{t J_a},k_{ t J_b}\}^2\,
    \end{equation}
    is satisfied, where $\vec\theta_i=\vec{k}_{ti}/E_i$.  If so, recombine
    $k_{J_a}$ and $k_{J_b}$ by adding transverse momenta vectorially, and setting
    the rapidity fraction of the resulting pseudo-particle $k_J$ to $\xi_J^{(\ell)}\simeq
    \xi_{J_a}^{(\ell)}\simeq \xi_{J_b}^{(\ell)}$.
  \end{itemize}
  We denote by $\ysc{k_1,\dots,k_n}$ the resulting value of $\ythree$,
  to distinguish it from $\yscbar{k_1,\dots,k_n}$ used to compute
  $\FNLL$.  Both algorithms have to be employed to compute the functions
  $\delta \mathcal{F}_{\rm clust}$ and $\delta \mathcal{F}_{\rm
    correl}$.

\item {\bf Cambridge algorithm:} in the case of the Cambridge
  algorithm, no recombination occurs whenever emissions are widely
  separated in angle. Therefore the clustering correction simply
  reduces to a clustering of two independently-emitted soft-collinear
  partons. In eq.~\eqref{eq:F1-clustering}, one can make the usual replacement
\begin{equation}
  \label{eq:y3-cam}
  \Theta\left(1 - \lim_{\ycut\to 0}\frac{\ysc{k_a,k_b,k_1,\dots,k_n}}{\ycut}\right)=
  \Theta\left(1-\lim_{\ycut \to 0}\frac{\ysc{k_a,k_b}}{\ycut} \right)\prod_{i=1}^n  \Theta\left(1-\lim_{\ycut\to 0}\frac{\ysc{k_i}}{\ycut}\right)\,,
\end{equation}
and observe that the contribution of any number of widely separated
emissions gives one, due to the normalization property of the measure
$\dZ$~\eqref{eq:dZ-norm}. As a consequence, the expression in
eq.~(\ref{eq:F1-clustering}) simplifies to
\begin{equation}
  \label{eq:dF-clust-cam}
  \begin{split}
    \delta\mathcal{F}_{\rm clust}(\lambda)&=
    \frac{1}{2!}\int_0^{\infty}\frac{d\zeta_a}{\zeta_a}\int_{0}^{2\pi}\frac{d\phi_a}{2\pi}\sum_{\ell_a=1,2}\int_0^1
    d\xi_a^{(\ell_a)}\left(\frac{2
        C_F\lambda}{\beta_0}\frac{R^{''}_{\ell_a}(\ycut)}{\alpha_s(\mu_R)}\right)
    \int_0^{\infty} \frac{d\kappa}{\kappa}
    \int_{-\infty}^{\infty}\!\!\! d\eta
    \int_{0}^{2\pi}\frac{d\phi}{2\pi} \times\\& \times
    \left[\Theta\left(1-\lim_{\ycut\to
          0}\frac{\ysc{k_a,k_b}}{\ycut}\right)-\Theta\left(1-\lim_{\ycut\to
          0}\frac{\yscbar{k_a,k_b}}{\ycut}\right)\right]\,,
  \end{split}
\end{equation}
which is non-zero only if the two emissions are clustered by the NNLL
algorithm, yielding
\begin{equation}
\begin{split}
  \delta\mathcal{F}_{\rm clust}(\lambda)&=
  \frac{1}{2!}\int_0^{\infty}\frac{d\zeta_a}{\zeta_a}\int_0^{2\pi}\frac{d\phi_a}{2\pi}\sum_{\ell_a=1,2}\int_0^1
  d\xi_a^{(\ell_a)}\left(\frac{2
      C_F\lambda}{\beta_0}\frac{R^{''}_{\ell_a}(\ycut)}{\alpha_s(\mu_R)}\right)
  \int_0^{\infty} \frac{d\kappa}{\kappa} \int_{-\infty}^{\infty}\!\!\!
  d\eta \int_0^{2\pi}\frac{d\phi}{2\pi} \times\\& \times
  \left[\Theta\left(1-\lim_{\ycut\to
        0}\frac{\ysc{k_a+k_b}}{\ycut}\right)-\Theta\left(1-\lim_{\ycut\to
        0}\frac{\max(\ysc{k_a},\ysc{k_b})}{\ycut}\right)\right]\Theta_{\rm
    clust}\,,
\end{split}
\end{equation}
where $\Theta_{\rm clust}$ restricts the allowed phase space to the
region where the two emissions $k_a$ and $k_b$ cluster. Using the
ordering variable for the Cambridge algorithm (eq.~\eqref{eq:vijC}) in
the small-angle approximation, emissions $a$ and $b$ will cluster if
\begin{equation}
\begin{split}
\label{eq:theta-clust-NNLL}
\vert\vec{\theta_{a}} - \vec{\theta_{b}}\vert^2 &<
\min\left\{\theta_{a},\theta_{b}\right\}^2 
\qquad
\Leftrightarrow
\qquad
\Theta_{\rm clust} = \Theta\left(\ln(2\cos\phi)-|\eta|\right)\Theta\left(\frac{\pi}{3}-|\phi|\right)\,.
\end{split}
\end{equation}

Applying these constraints gives
\begin{equation}
\begin{split}
    \delta\mathcal{F}_{\rm clust}(\lambda)&=
 \sum_{\ell_a=1,2}\left(\frac{2C_F\lambda}{\beta_0}\frac{R^{''}_{\ell_a}(\ycut)}{\alpha_s(\mu_R)}\right)
    \int_0^{\infty} \frac{d\kappa}{\kappa}
    \int_{-\frac{\pi}{3}}^{\frac{\pi}{3}}\frac{d\phi}{2\pi} \ln(2\cos\phi)
    \ln\left(\frac{\max\left\{1,\kappa^2\right\}}{1+\kappa^2+2\kappa\cos\phi}\right)
    \\
&\approx \sum_{\ell_a=1,2}\left(\frac{2C_F\lambda}{\beta_0}\frac{R^{''}_{\ell_a}(\ycut)}{\alpha_s(\mu_R)}\right)(-0.493943)\,.
  \end{split}
\end{equation}
\end{itemize}

\subsubsection{Correlated corrections}
The correlated correction describes an ensemble of
independently-emitted soft-collinear partons of which one branches
into either a quark or a gluon pair, denoted by $k_a$ and $k_b$. The
property of rIRC safety ensures that the splitting can be treated
inclusively at NLL, and at this order it contributes to the Sudakov
radiator~\cite{Banfi:2014sua}. At NNLL the splitting must be resolved,
and this is taken into account by correcting the inclusive
approximation. This leads to
\begin{equation}
  \label{eq:F1-correl}
  \begin{split}
    \delta\mathcal{F}_{\rm
      correl}(\lambda)&=\int_0^{\infty}\frac{d\zeta_a}{\zeta_a}\int_0^{2\pi}\frac{d\phi_a}{2\pi}\sum_{\ell_a=1,2}\int_0^1
    d\xi_a^{(\ell_a)}\left(\frac{2
        C_F\lambda}{\beta_0}\frac{R^{''}_{\ell_a}(\ycut)}{\alpha_s(\mu_R)}\right)
    \int_0^{\infty} \frac{d\kappa}{\kappa}
    \int_{-\infty}^{\infty}\!\!\! d\eta
    \int_0^{2\pi}\frac{d\phi}{2\pi} \frac{1}{2!}
    C_{ab}(\kappa,\eta,\phi) \times\\& \times \int \dZ
    \left[\Theta\left(1-\lim_{\ycut\to
          0}\frac{\ysc{k_a,k_b,\{k_i\}}}{\ycut}\right)-\Theta\left(1-\lim_{\ycut\to
          0}\frac{\yscbar{k_a+k_b,\{k_i\}}}{\ycut}\right)\right]\,,
  \end{split}
\end{equation}
where
\begin{equation}
C_{ab}(\kappa,\eta,\phi) = \frac{\tilde M^2(k_a,k_b)}{M_{sc}^2(k_a)M_{sc}^2(k_b)}\,,
\end{equation}
is the ratio of the correlated soft matrix element
$\tilde M^2(k_a,k_b) = M^2(k_a,k_b) -M^2(k_a)M^2(k_b)$ (i.e.~the
difference between the full two-parton matrix element in the soft
limit and the independent emission contribution) to the product of the
two independent soft-collinear matrix elements for the emissions $k_a$
and $k_b$. Notice that $C_{ab}$ depends only on the correlation
variables $\kappa,\eta,\phi$ defined in eq.~\eqref{eq:corvars}.
The observable $\ysc{k_a,k_b,\{k_i\}}$ is computed with the same
algorithm used for the clustering correction. In the inclusive
approximation $\ysc{k_a+k_b,\{k_i\}}$ reduces to the NLL value
$\yscbar{k_a+k_b,\{k_i\}}$. As done for the clustering correction, we
impose that $k_b$ belongs to the same hemisphere as $k_a$ in order to
neglect undesired subleading effects. While for the Durham the
observable $\ysc{k_a,k_b,\{k_i\}}$ is computed using the algorithm
given above for the clustering corrections, in the case of the
Cambridge the final expression simplifies considerably.

\begin{itemize}
\item {\bf Cambridge algorithm:} in the case of the Cambridge
  algorithm we can integrate out the harmless rapidity-separated
  soft-collinear ensemble and write
\begin{equation}
  \label{eq:dF-correl-cambridge}
  \begin{split}
    \delta\mathcal{F}_{\rm
      correl}(\lambda)&=\int_0^{\infty}\frac{d\zeta_a}{\zeta_a}\int_0^{2\pi}\frac{d\phi_a}{2\pi}\sum_{\ell_a=1,2}\int_0^1
    d\xi_a^{(\ell_a)}\left(\frac{2C_F\lambda}{\beta_0}\frac{R^{''}_{\ell_a}(\ycut)}{\alpha_s(\mu_R)}\right)
    \int_0^{\infty} \frac{d\kappa}{\kappa}
    \int_{-\infty}^{\infty}\!\!\! d\eta
    \int_0^{2\pi}\frac{d\phi}{2\pi} \frac{1}{2!}
    C_{ab}(\kappa,\eta,\phi) \times\\& \times
    \left[\Theta\left(1-\lim_{\ycut\to
          0}\frac{\max\left\{\ysc{k_a},\ysc{k_b}\right\}}{\ycut}\right)-\Theta\left(1-\lim_{\ycut\to
          0}\frac{\ysc{k_a+k_b}}{\ycut}\right)\right](1-\Theta_{\rm clust})\,,
  \end{split}
\end{equation}
where $\Theta_{\rm clust}$ is defined in eq.~\eqref{eq:theta-clust-NNLL}.
It is clear in fact that $\delta\cal{F}_{\rm correl}$ is non-zero only when
$k_a$ and $k_b$ are not clustered in the first $\Theta$ function.
This leads to
\begin{equation}
\begin{split}
  \delta\mathcal{F}_{\rm
    correl}(\lambda)
  & =
  \sum_{\ell_a=1,2}\left(\frac{2C_F\lambda}{\beta_0}\frac{R^{''}_{\ell_a}(\ycut)}{\alpha_s(\mu_R)}\right)
  \int_0^{\infty}
  \frac{d\kappa}{\kappa}\int_{0}^{2\pi}\frac{d\phi}{2\pi}
  \frac{1}{2!}  \int_{-\infty}^{\infty}\!\!\! d\eta\,
  C_{ab}(\kappa,\eta,\phi)\ln\left(\frac{1+\kappa^2+2\kappa\cos\phi}{\max\left\{1,\kappa^2\right\}}\right)(1-\Theta_{\rm clust})\,.\\
\end{split}
\end{equation}

This integral can be evaluated numerically giving
\begin{equation}
\begin{split}
  \delta\mathcal{F}_{\rm correl}(\lambda)
  &\approx\sum_{\ell_a=1,2}\left(\frac{\lambda}{\beta_0}\frac{R^{''}_{\ell_a}(\ycut)}{\alpha_s(\mu_R)}\right)(2.1011(2)
  C_A + 1.496(1)\times10^{-2} n_f)\,,
\end{split}
\end{equation}
where the number in round brackets is the uncertainty in the last digit.
\end{itemize}

\subsubsection{Hard-collinear and recoil corrections}

The hard-collinear and recoil corrections describe configurations in
which a parton of the ensemble is emitted collinearly to one of the
Born legs and carries a significant fraction $z$ of the emitter's momentum. 

The hard-collinear correction takes into account the exact matrix
element for this hard-collinear emission:
\begin{equation}
  \label{eq:Fhc}
  \begin{split}
&  \delta\mathcal{F}_{\rm hc}(\lambda)=\sum_{\ell=1,2}
\frac{\alpha_s(\sqrt{\ycut}Q)}{2\alpha_s(\mu_R)}
   \int_0^{\infty}\frac{d\zeta}{\zeta}
   \int_{0}^{2\pi}\frac{d\phi}{2\pi} \int \dZ
   \times \\ & \times
   \int_0^1 \!\frac{dz}{z}\,(z p_{qg}(z) - 2 C_F )
   \left[\Theta\left(1-\lim_{\ycut\to 0} \frac{\yscbar{k, \{k_i\}}}{\ycut}\right)
     -\Theta\left(1-\lim_{\ycut\to 0} \frac{\yscbar{\{k_i\}}}{\ycut}\right)\Theta(1-\zeta)\right]\,,
  \end{split}
\end{equation}
where $z p_{qg} = C_F(1+(1-z)^2)$. 

Similarly, the recoil correction implements the effect of the
hard-collinear emission on the observable by taking into account the
exact recoil kinematics in this regime. 
In fact, for a hard-collinear parton $k$, the approximation
$\yscbar{k}=(k_t/Q)^2$ is no longer valid. In order to compute
$\yscbar{k}$ it is convenient to express the final transverse momenta
with respect to the thrust axis of the event. In the above
configuration, we have to distinguish between the transverse momentum
$k_t$ of $k$ with respect to the final
thrust axis and its transverse momentum $k_t'$ with respect to the
emitter prior to the hard-collinear emission, as discussed in
ref.~\cite{Banfi:2014sua}.  In turn this will give the correct
transverse momentum of the collinear emission with respect to the
final direction of the emitter, which enters the definition of
$\ythree$. Denoting by $z$ the momentum fraction carried away from the
emitter by the hard collinear parton $k$ emitted off the Born leg
$\tilde{p}_\ell$, we have~\cite{Banfi:2014sua}
\begin{equation}
\label{eq:kthc}
\vec k_t\simeq\vec k'_t +z \vec{p}^{\,\prime}_{\ell, t} \,,\qquad \vec{p}^{\,\prime}_{\ell, t}\equiv -\sum_{i\,\in\,{\cal H}^{(\ell)}}\vec
k_{ti}\,,
\end{equation}
where the sum runs over all of the remaining soft-collinear emissions
emitted off $\tilde{p}_\ell$, for which $z_i\to 0$ (for these
emissions the transverse momentum w.r.t. the thrust axis coincides
with the one computed w.r.t. the emitter). The corresponding
expression for the $\yhc{k}$ becomes
\begin{equation}
\label{eq:y3hc}
\begin{split}
  \yhc{k}=& \frac{\min\left\{z,1-z\right\}^2}{Q^2}\left|\frac{\vec{k}_{t}}{z} -
            \frac{\vec{p}_{\ell, t}}{1-z}\right|^2
          = \min\left\{\frac{1}{1-z},\frac{1}{z}\right\}^2 \left(\frac{k'_t}{Q}\right)^2\,,
\end{split}
\end{equation}
where $\vec{p}_{\ell, t}=\vec{p}^{\,\prime}_{\ell, t}-\vec{k}_{t}$ is
the transverse momentum of the Born emitter $\tilde{p}_\ell$ with
respect to the thrust axis.  Note that, since $k$ is the most
energetic parton of the ensemble, its rapidity fraction is by
construction the largest of all given that all transverse momenta are
of the same order in virtue of rIRC safety.
The recoil correction then takes the form~\cite{Banfi:2014sua}
\begin{equation}
  \label{eq:Frec}
  \begin{split}
  \delta\mathcal{F}_{\rm rec}(\lambda)&=\sum_{\ell=1,2}
   \frac{\alpha_s(\sqrt{\ycut}Q)}{2\alpha_s(\mu_R)}
   \int_0^{\infty}\frac{d\zeta}{\zeta}
   \int_{0}^{2\pi}\frac{d\phi}{2\pi}\int \dZ
     \times \\ & \times
   \int_0^1 \!dz\,p_{qg}(z)
   \left[\Theta\left(1-\lim_{\ycut\to 0} \frac{\yhc{k',\{k_i\}}}{\ycut}\right)
     -\Theta\left(1-\lim_{\ycut\to 0} \frac{\yscbar{k,\{k_i\}}}{\ycut}\right)\right]\,,
  \end{split}
\end{equation}
where $\zeta \ycut  Q^2 =(k_t')^2 $, and  the momentum of the hard-collinear gluon $k'$ is a function of
$\zeta$, $\vec{{p}}_{\ell, t}' $, and $z$. The momentum $k$ in the second theta-function
is obtained from $k'$ by taking the limit $z\to 0$. The observable
$\yhc{k',\{k_i\}}$ appearing in the first theta-function is
computed in the hard-collinear limit by means of the algorithms
defined below.
\begin{itemize}
\item {\bf Durham algorithm:}  in the considered kinematic configuration, the
  NLL algorithm is modified as follows:
\begin{itemize}
\item[1.] Find the index $I$ of the parton with the smallest
  $\yfun{k_i}(=\yscbar{k_i}$ for the soft-collinear partons and
  $\yhc{k}$ of eq.~\eqref{eq:y3hc} for the hard-collinear one).
\item[2.] Find $k_J$ as in step 2 of the NLL algorithm.
\item[3.] If $k_J$ is found, recombine partons $I$ and $J$ into a new
  pseudo-particle $k_P$ with
  $\vec{k}_{tP}=\vec{k}_{tI}+\vec{k}_{tJ}$ and
  $\xi_P^{(\ell)}=\xi_J^{(\ell)}$. Otherwise, $k_I$ is clustered with
  the Born leg $\tilde{p}_\ell$ it was emitted off as
  $\vec{p}_{\ell, t}=\vec{k}_{tI}+\vec{p}_{\ell, t}$, and
  removed from the list of pseudo-particles. If $k_P$ contains the
  hard-collinear parton (say parton $k_I=k$ is the hard-collinear one)
  the corresponding $\yfun{k_{P}}$ will be
  \begin{equation}
    \yhc{k_P} =
 \frac{\min\left\{z,1-z\right\}^2}{Q^2}\left|\frac{\vec{k}_{tP}}{z} -
            \frac{\vec{p}_{\ell, t}}{1-z}\right|^2\notag\,.
  \end{equation}
 This quantity will be used in step 1 of the next iteration.
\item[4.]  Repeat until only one pseudo-particle $k_P$
  remains, and set $\yfun{k_1,\dots,k_n}=\yfun{k_P}$.
\end{itemize}
This algorithm is applied to the computation of the observable $\yhc{k',\{k_i\}}$.
\item {\bf Cambridge algorithm:}  for the Cambridge algorithm, one can
  perform the same replacements used for the other corrections above,
  and notice that the measure $\dZ$ integrates to one. We then have
\begin{equation}
  \label{eq:dFhc-cambridge}
  \begin{split}
&  \delta\mathcal{F}_{\rm  hc}(\lambda)=\sum_{\ell=1,2}
\frac{\alpha_s(\sqrt{\ycut}Q)}{2\alpha_s(\mu_R)}
   \int_0^{\infty}\frac{d\zeta}{\zeta}
   \int_0^{2\pi}\frac{d\phi}{2\pi}
   \times \\ & \times
   \int_0^1 \!\frac{dz}{z}\,(z p_{qg}(z) - 2 C_F )
   \left[\Theta\left(1-\lim_{\ycut\to 0} \frac{\yscbar{k}}{\ycut}\right)
     -\Theta(1-\zeta)\right]=0\,, 
  \end{split}
\end{equation}
where in the last step we used the definition of $\zeta = \ysc{\{k\}}/\ycut$.

For the recoil correction, the same argument leads to a simplified
formula where the contribution from the hard-collinear emission
factorizes with respect to the soft-collinear ones. Since the
hard-collinear emission $k'$ propagates at very high rapidity, it is
widely separated in rapidity from the soft-collinear ensemble. This
leads to
\begin{equation}
  \label{eq:theta-hc-cambridge}
  \Theta\left(1-\lim_{\ycut\to 0}
    \frac{\yhc{k',\{k_1,\dots,k_n\}}}{\ycut}\right) = \Theta\left(1-\lim_{\ycut\to 0}
    \frac{\yhc{k'}}{\ycut}\right) \prod_{i=1}^n \Theta\left(1-\lim_{\ycut \to 0} \frac{\ysc{k_i}}{\ycut}\right)\,,
\end{equation}
which shows that the recoil correction is non-zero only if
$\yhc{k',\{k_i\}} > \ysc{\{k_i\}}$ for all $i$. With this condition
one obtains
\begin{equation}
  \label{eq:dFrec-cambridge}
  \begin{split}
  \delta\mathcal{F}_{\rm rec}(\lambda)&=\sum_{\ell=1,2}
   \frac{\alpha_s(\sqrt{\ycut}Q)}{2\alpha_s(\mu_R)}
   \int_0^{\infty}\frac{d\zeta}{\zeta}
   \int_0^{2\pi}\frac{d\phi}{2\pi}
     \times \\ & \times
   \int_0^1 \!dz\,p_{qg}(z)
   \left[\Theta\left(1-\lim_{\ycut\to 0} \frac{\yhc{k'}}{\ycut}\right)
     -\Theta\left(1-\lim_{\ycut \to 0} \frac{\yscbar{k}}{\ycut}\right)\right]\,.
  \end{split}
\end{equation}
Using eq.~(\ref{eq:y3hc}) one gets
\begin{equation}
  \label{eq:y3-hc-k}
 \frac{\yhc{k'}}{\ycut} = \min\left\{\frac{1}{1-z},\frac{1}{z}\right\}^2\zeta=\frac{1}{\max(z^2,(1-z)^2)}\zeta\,,
\end{equation}
which can be plugged in eq.~\eqref{eq:dFrec-cambridge} to obtain
\begin{equation}
  \label{eq:dFrec-cambridge-final}
  \begin{split}
    \delta\mathcal{F}_{\rm rec}(\lambda)&=
    \frac{\alpha_s(\sqrt{\ycut}Q)}{\alpha_s(\mu_R)} \int_0^1
    \!dz\,p_{qg}(z) \ln\left[\max(z^2,(1-z)^2)\right]=C_F
    \frac{\alpha_s(\sqrt{\ycut}Q)}{\alpha_s(\mu_R)}
    \left(3-\frac{\pi^2}{3}-3\ln 2\right)\,.
  \end{split}
\end{equation}
\end{itemize}

\subsubsection{Soft-wide-angle corrections}
This correction describes the contribution from configurations where
an ensemble of soft-collinear partons is accompanied by a soft
emission $k$ at wide angles with respect to the hard legs. It takes
the form~\cite{Banfi:2014sua}
\begin{equation}
  \label{eq:dF-wa}
\begin{split}
 \delta\mathcal{F}_{\rm wa}(\lambda) &= C_F \frac{\alpha_s(\sqrt{\ycut}
   Q)}{\alpha_s(\mu_R)} \int_0^{\infty} \frac{d\zeta}{\zeta}
 \int_{-\infty}^{\infty} \!\! d\eta \int_{0}^{2\pi}\! \frac{d\phi}{2\pi}\int \dZ
 \\ & \times
\left[\Theta\left(1-\lim_{\ycut \to 0}\frac{\ywa{k, \{k_i\}}}{\ycut}\right)-
\Theta\left(1-\lim_{\ycut \to 0}\frac{\yscbar{k,\{k_i\}}}{\ycut}\right)\right]\,,
\end{split}
\end{equation}
with $\zeta \ycut Q^2 = k_t^2$, and $\eta$ the emission's rapidity
with respect to the thrust axis.  The observable $\ywa{k, \{k_i\}}$ in
the soft, wide-angle configuration can be computed as described below
for the various algorithms.
\begin{itemize}
\item {\bf Durham algorithm:} since, by definition, the wide-angle
  emission has the smallest rapidity fraction amongst all emissions,
  if it recombines with any of the other collinear partons, it will be
  pulled at larger rapidity fractions (see step 3 of the NLL
  algorithm). Therefore, the result of the recombination will be the
  same as if $k$ were soft and collinear. It follows that the
  soft-wide angle contribution is non-zero only if $k$ does not
  cluster with any of the soft-collinear emissions. For this emission,
  the expression of $\ythree$ becomes
\begin{equation}
\label{eq:yijDwa}
\yfun{k}=2\frac{E^2}{Q^2} \left( 1-|\cos\theta|\right)\,,
\end{equation}
where $\theta$ is the angle with respect to the direction identified
by the Born momenta, which remain back-to-back in the presence of soft
emissions, and practically coincides with the thrust axis. The
corresponding observable $\ywa{k, \{k_i\}}$ can be computed by means
of the NLL algorithm for strongly-ordered emissions, where one uses
eq.~\eqref{eq:yijDwa} to express $\yfun{k}$ for the soft-wide-angle
emission $k$. As soon as the latter is clustered with any of the
remaining soft-collinear emissions, the algorithm simply reduces to
the NLL one in its original form.
\item {\bf Cambrige algorithm:}  
since the Cambridge algorithm does not cluster objects widely separated
in rapidity, in this case the only non-trivial contribution comes when
the soft wide-angle emission is the last particle to be recombined,
namely if $\ywa{k} > \yscbar{ \{k_i\} }$.
  We then obtain
\begin{equation}
  \label{eq:dF-soft-simp-final}
\begin{split}
 \delta\mathcal{F}_{\rm wa}(\lambda) &= C_F\frac{\alpha_s(\sqrt{\ycut}
   Q)}{\alpha_s(\mu_R)} \int_0^{\infty} \frac{d\zeta}{\zeta}
 \int_{-\infty}^{\infty} \!\! d\eta \int_{0}^{2\pi}\! \frac{d\phi}{2\pi}\int \dZ
 \\ & \times
\left[\Theta\left(1-\lim_{\ycut \to 0}\frac{\ywa{k}}{\ycut}\right)-\Theta\left(1-\lim_{\ycut \to 0}\frac{\ysc{k}}{\ycut}\right)\right]\,.
\end{split}
\end{equation}
Rephrasing eq.~\eqref{eq:yijDwa} in terms of $\zeta$ and $\eta$, the
three-jet resolution parameter for emission $k$ is given by
\begin{equation}
  \label{eq:y3-soft}
  \frac{\ywa{k}}{\ycut} = \zeta \left(1+e^{-2|\eta|}\right)\,,
\end{equation}
from which it follows that
\begin{equation}
  \label{eq:dF-soft-final}
\begin{split}
  \delta\mathcal{F}_{\rm wa}(\lambda) = -C_F
  \frac{\alpha_s(\sqrt{\ycut} Q)}{\alpha_s(\mu_R)}
  \int_{-\infty}^{\infty} \!\!\! d\eta \,
  \ln\left(1+e^{-2|\eta|}\right)=-C_F
  \frac{\pi^2}{12}\frac{\alpha_s(\sqrt{\ycut} Q)}{\alpha_s(\mu_R)}\,.
\end{split}
\end{equation}
\end{itemize}
\newpage
\subsection{Check of logarithmic expansion against the
  ${\cal O}(\alpha_s^2)$ fixed-order prediction.}
\begin{figure}[t]
  \centering
    \subfigure{
  \includegraphics[width=0.48\linewidth]{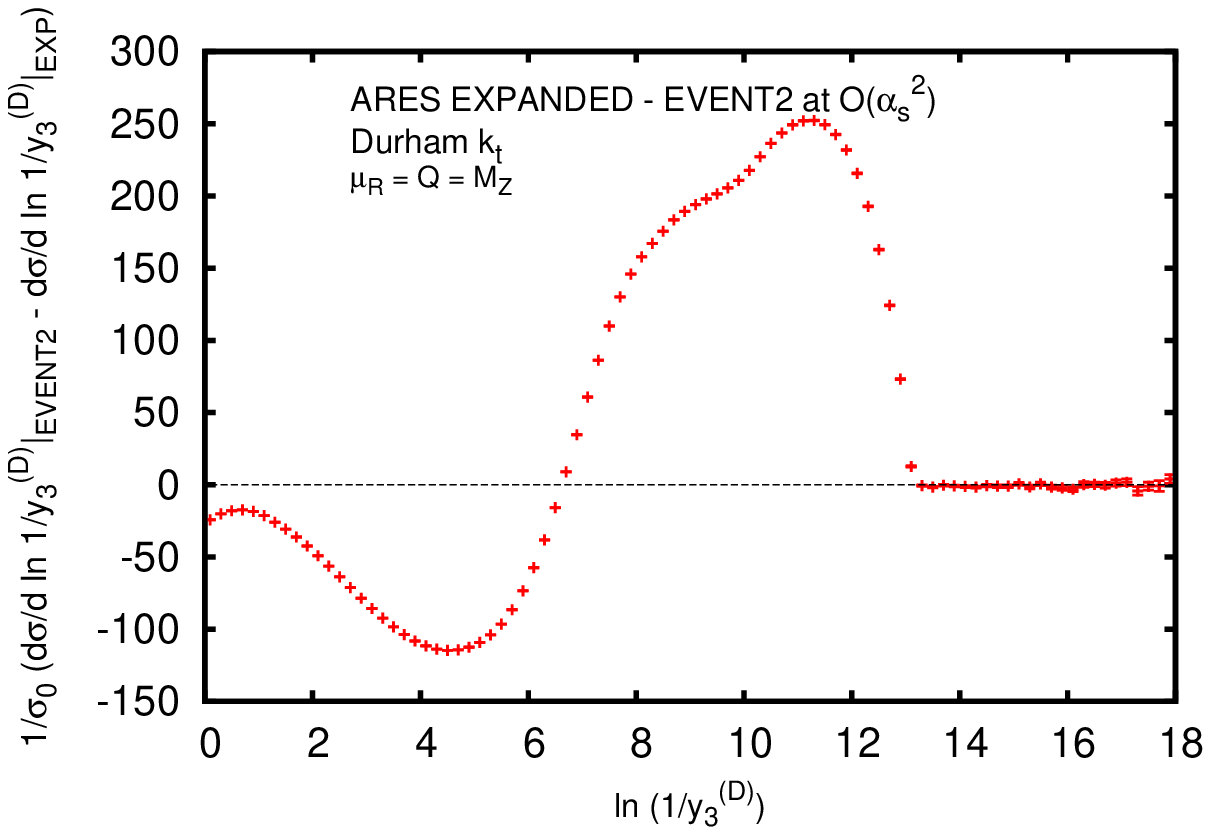}}
  \subfigure{
  \includegraphics[width=0.48\linewidth]{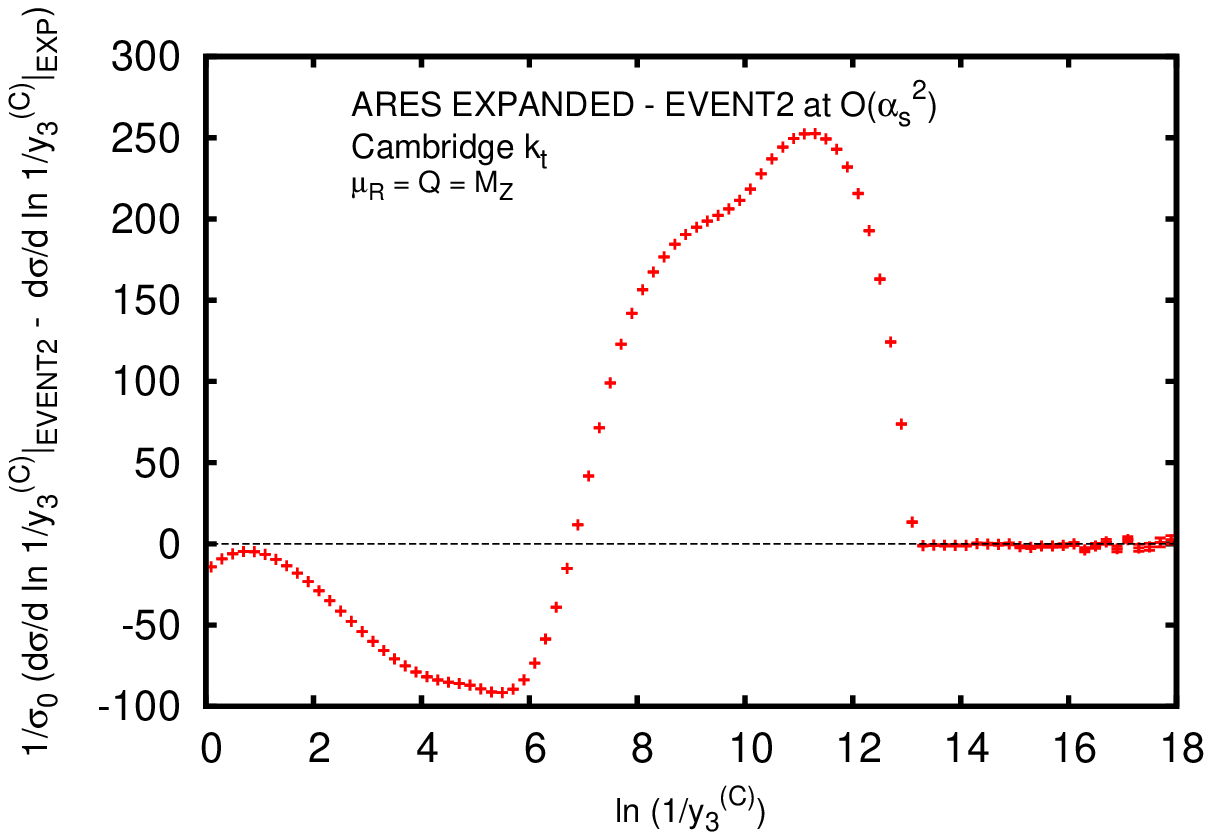}}
 \caption{Difference of the three-jet resolution differential
   distributions between  {\tt Event2}  and the ${\cal O}(\alpha_s^2)$ expansion
 of the resummation performed in this letter.}
\label{fig:check-logs-single}
\end{figure}

In this section we report on the check of the resummation formula by comparing its expansion to ${\cal O}(\alpha_s^2)$ to the exact fixed-order result provided by the generator \texttt{EVENT2}~\cite{Catani:1996vz}.  
In particular, we compare the normalized differential distributions
\begin{equation}
\frac{1}{\sigma_0} \frac{d\sigma}{d \ln (1/y_{3})}\,,
\end{equation}
with $\sigma_0$ being the Born cross section for $e^+e^-\to 2$\,jets.
Figure~\ref{fig:check-logs-single} shows the comparison for the Durham and the Cambridge algorithm. As expected from a NNLL result, the difference between the two predictions approaches zero for asymptotically large values of $\ln (1/\ythree)$.

\begin{figure}[htbp]
  \centering
  \includegraphics[width=0.465\linewidth]{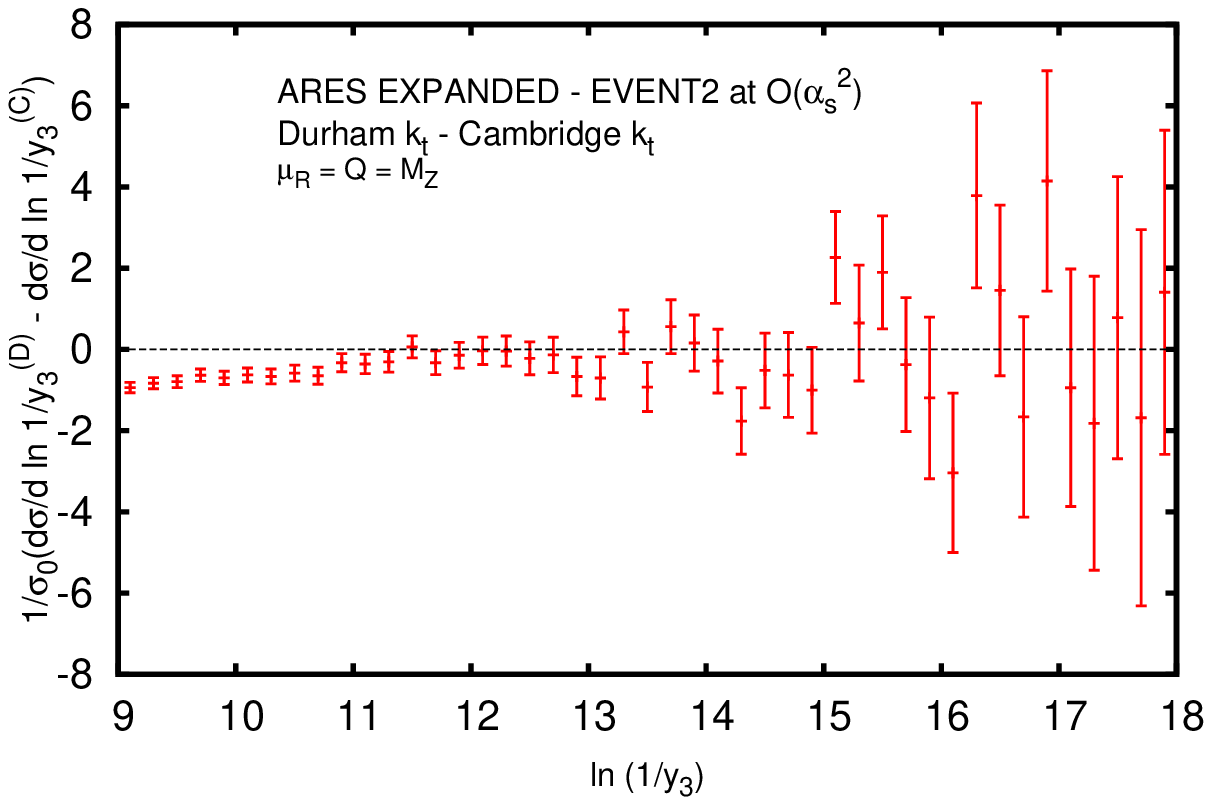}
 \caption{Difference of the three-jet resolution differential
   distribution between the Durham and the Cambridge algorithm as
 computed in {\tt Event2} minus the ${\cal O}(\alpha_s^2)$ expansion
 of the resummation performed in this letter.}
\label{fig:check-logs}
\end{figure}

Since the two observables in this case are identical for a single emission, it is useful to perform a similar check on the difference
\begin{equation}
\label{eq:Deltasigma}
\frac{1}{\sigma_0}\left(\frac{d\sigma}{d \ln (1/y_{3}^{(D)})} - \frac{d\sigma}{d \ln (1/y_{3}^{(C)})}\right)\,.
\end{equation}
Taking the difference in eq.~\eqref{eq:Deltasigma} leads to numerically more stable fixed-order results. The corresponding check is shown in fig.~\ref{fig:check-logs}.

\end{document}